\pdfoutput=1

\documentclass[pra,10pt,twocolumn,superscriptaddress, footnoteinbib]{revtex4}
\usepackage{amsmath}
\usepackage{latexsym}
\usepackage{amssymb}
\usepackage{graphics,epstopdf}
\usepackage[colorlinks=true, citecolor=blue, urlcolor=blue ]{hyperref}
\usepackage{epsf,graphics,graphicx}

\newcommand{\n}{\noindent}

\newcommand{\ed}{\end{document}}
\newcommand{\beq}{\begin{equation}}
\newcommand{\eeq}{\end{equation}}

\begin{document}
\title{The Geometric phase and fractional  orbital angular momentum states in electron vortex beams 
}
\author{Pratul Bandyopadhyay}\email{b$_$pratul@yahoo.co.in, retired Prof.}
\author{Banasri Basu}\email{sribbasu@gmail.com}
\affiliation{Physics and Applied Mathematics Unit, Indian Statistical Institute, Kolkata 700108, India}
\author{Debashree Chowdhury}\email{debashreephys@gmail.com, Present address: Department of Physics,Ben-Gurion University,Beer Sheva 84105,Israel}
\affiliation{Department of Physics, ​Harish-Chandra Research institute,
Chhatnag Road, Jhusi, Allahabad, U. P. 211019,India}


\begin{abstract}
\n
We study here fractional orbital angular momentum (OAM) states in electron vortex beams (EVB) from the perspective of geometric phase. We have considered the skyrmionic model of an electron, where it is depicted as a scalar electron orbiting around the vortex line, which gives rise to the spin degrees of freedom. The geometric phase acquired by the scalar electron orbiting around the vortex line induces the  spin-orbit interaction. This leads to the fractional OAM states when we have non-quantized monopole charge associated with the corresponding geometric phase. This involves tilted vortex in EVBs. The monopole charge undergoes the renormalization group (RG) flow, which incorporates a length scale dependence making the fractional OAM states unstable upon propagation. It is pointed out that when EVBs move in an external magnetic field, the Gouy phase associated with the Laguerre-Gaussian modes modifies the geometric phase factor and a proper choice of the radial index helps to have a stable fractional OAM state.

\end{abstract}

\maketitle
 \section{Introduction}
 It is now well known that optical vortex beams \cite{1} carrying orbital angular momentum(OAM) exhibit an azimuthal phase structure $exp(i \ell\phi),$ where $\ell$ is an integer number implying OAM of $\ell\hbar$ per photon \cite{allen}. These can be produced in the laboratory using spiral phase plates \cite{2} or computer generated holograms \cite{3,opticslett}. Such optical beams are the Laguerre-Gaussian(LG) modes, which imprint a $2\pi l$ step in the phase of the electromagnetic field. However, it is also possible to generate a situation such that the phase step is not an integer multiple of $2\pi.$ This corresponds to fractional OAM. 
 Light emerging from a fractional phase step in general is unstable on propagation. 
 Gotte et. al. \cite{7} have pointed out that fractional OAM can be produced through a generic superposition of light modes with different values of $\ell$ and LG beams with a minimal number of different Gouy phases increase propagational stability. These states can be decomposed into basis of integer OAM states.
 
 In recent times electron vortex beams(EVBs) with OAM have been produced experimentally \cite{9,10,11}. An EVB is generally visualized as a scalar electron orbiting around the vortex line. Bliokh et. al. \cite{12} studied the relativistic EVBs representing the angular momentum eigenstate of a free Dirac equation and constructed exact Bessel beam solution.  In a recent paper \cite{13}, it has been argued that in the skyrmionic model of an electron, where it is depicted  as a scalar electron rotating around a direction vector(vortex line), which is topologically equivalent to a magnetic flux line corresponding to the spin degrees of freedom, EVBs appear as a natural consequence. Evidently, the vortex here is a spin vortex. Apart from this, when the orbiting electron carries OAM, EVBs exhibit an azimuthal phase structure $exp(i \ell\phi).$
 
 It has been observed that optical and electron vortex beams carrying OAM share much similarities in their behaviour. Indeed, just like EVBs, the dynamics of the optical vortex beams(OVBs) can also be studied from the perspective of geometric phase acquired by the beam field orbiting around the vortex line \cite{16}. In this note, we shall study the situation when EVBs carry fractional OAM. When the scalar electron in an EVB orbiting around the vortex line, acquires the Berry phase, which involves quantized monopole charge, we have screw or edge dislocation \cite{13,16}. The screw dislocation essentially corresponds to paraxial beams, where the vortex line is parallel to the wave front propagation direction, whereas for edge dislocation, the vortex line is orthogonal to the wave front propagation direction. Besides, there are situations when we have tilted vortex, where the vortex line makes an angle $\theta$ with respect to the wave front propagation direction such that $0< \theta< \pi/2,$  which corresponds to the mixed screw-edge dislocation. In this case the associated Berry phase involves non-quantized monopole charge and incorporates spin-orbit interaction(SOI), which effectively induces fractional OAM. It is here argued that for EVBs the structural stability arises, when these beams move in an external magnetic field, where we can have LG modes such that the monopole charge of the Berry phase is modified by the Gouy phase\cite{18}. This analysis suggests that fractional OAM in vortex beams is a natural phenomenon and can be visualized properly from the perspective of the Berry phase.


 In sec. 2, we shall recapitulate certain features of the Berry phase in EVBs and its association with  the spin-orbit interaction leading to fractional OAM states.  In sec. 3, we shall discuss the fractional OAM states from the view point of the monopole harmonics. 
 In sec. 4 we shall consider the stability of such beams when these move in an external magnetic field and the relevant consequences.

\section{Electron Vortex beams with fractional orbital angular momentum and the Berry phase}
 
In a recent paper\cite{13}, the geometrodynamics of EVBs have been analysed from the point of view of  the geometric phase. To this end, we have considered the skyrmionic model of an electron, where it is depicted  as a scalar electron orbiting around a direction vector(vortex line), which gives rise to the spin degrees of freedom \cite{14}. Evidently, the scalar electron moving around the vortex line acquires the Berry phase after forming a closed loop. It has been pointed out that for paraxial beams, the Berry phase is vanishing. For non-paraxial beams, corresponding to the edge dislocation, where the vortex line is orthogonal to the wave front propagation direction, the Berry phase involves quantized monopole. However, for tilted vortex, where the vortex line makes an arbitrary angle with the wave-front propagation direction, the corresponding Berry phase involves non-quantized monopole charge. It is here noted that in this case electron carries fractional OAM. The situation involves spin-orbit interaction(SOI) and fractional OAM appears as a consequence of this. 

The Berry phase acquired by the scalar electron encircling the vortex line is $2\pi\mu$ \cite{24}, where $\mu$ is the monopole charge. In terms of the  solid angle subtended by the closed circuit at the origin of the unit sphere, where the monopole is located, the Berry phase is given by $ \mu\Omega(C),$  where $\Omega(C)$ is the solid angle given by 
\beq \Omega(C) = \int_{C}(1 - cos \theta)d\phi = 2\pi(1- cos\theta).\label{eq8}\eeq
Here $\theta$ is the polar angle of the vortex line with the quantization axis(z axis) and $\mu$ corresponds to the monopole charge, which effectively represents the spin. Indeed, the total angular momentum of a charged particle in the field of a magnetic monopole of charge  $\mu$ is given by $\vec{J} = \vec{L} - \mu\hat{\vec{r}},$ where $\vec{L}$ is the orbital angular momentum(OAM). In case, where OAM is vanishing,  $\mu$ corresponds to the total angular momentum of the particle, i.e. the spin angular momentum(SAM) with $s_{z} = \pm \mu$.  One may note that for $\mu =1/2,$  and the corresponding Berry phase derived from (1) is,
\beq  \phi_{B} = \pi(1 - cos\theta)\label{eq91}.\eeq

If one considers a reference frame where the scalar electron in the EVB is taken to be fixed and the vortex state moves in the field of a magnetic monopole around a closed path, evidently, $\phi_{B}$ in (\ref{eq91}) will  correspond to the Berry phase acquired by the vortex state(spin state) in an EVB. 
The angle $\theta$ represents the  deviation of the vortex line from the $z$ axis. 
Equating this phase  $\phi_{B}$ with  $2\pi\mu,$ which is the Berry phase acquired by the scalar electron moving around the vortex line in the closed path, we find that the effective monopole charge associated with the corresponding vortex line having  polar angle $\theta$ with the $z$ axis is given by
\beq  \mu = \frac{1}{2}(1-cos\theta)\label{9}.\eeq

Eqn. (3) suggests that  the paraxial beams with  $\theta =0$  and orthogonal non-paraxial beams with $\theta=\pi/2$ correspond to quantized monopole  charge.    However, for the continuous range  $0 < ~\theta<~\frac{\pi}{2},$ the corresponding monopole charge $\mu$ is non-quantized and involves tilted vortices. 
Non-paraxial beams in general incorporates SOI, which modifies the OAM $\langle\vec{L}\rangle$ as well as spin $\langle\vec{S}\rangle$. Introducing the mapping $\vec{L} = l\hat{z}$ and $\vec{S} = s\hat{z},$ we note that EVBs having tilted vortex, which are characterized by the Berry phase associated with the non-quantized monopole charge, the OAM as well as SAM is modified as
\beq \langle\vec{L}\rangle = (l -\mu)\hat{z} ~~~~~~~\langle\vec{S}\rangle = (s +\mu)\hat{z}.\eeq

This suggests that the OAM in such a situation can take any arbitrary fractional value, as the factor $\mu$ can take any value between $0$ to $1.$ It is observed that the specific case for the quantized fractional value $\mu = \frac{1}{2},$ corresponds to the non-paraxial beam, where the vortex is orthogonal to the wave-front propagation direction. In analogy with the central charge of conformal field theory, the monopole charge undergoes the renormalization group (RG) flow \cite{27,17}. When $\mu$ depends on a certain parameter $\lambda,$ for certain fixed points $\lambda^{*},$ $\mu$ takes quantized values. However, for other values of $\lambda,$ $\mu$ is non-quantized. The RG flow suggests that $\mu$ is a decreasing function such that $L\frac{\partial \mu}{\partial L} \leq 0,$ where $L$ is the length scale. Thus we observe that for non-quantized value of $\mu,$ it changes with a characteristic length scale. This implies that EVBs carrying fractional OAM will be unstable on propagation. However it will be shown here that, when EVBs move in an external magnetic field, we can attain the stability of these states. Indeed in this case the Berry phase factor $\mu$ is changed through the incorporation of the Gouy phase associated with the corresponding LG modes. The stability of the fractional OAM state is attained by choosing an appropriate value of the radial index, which essentially corresponds to the choice of a fixed value of the length scale $L.$ 

\section{Monopole harmonics and fractional orbital angular momentum}

It has been shown in some earlier papers \cite{16,Bayen,Sir} that the presence of the spin vector elevates the localization region of a relativistic massive as well as a massless particle from $S^{2}$ to $S^{3}.$ Indeed noting that $S^{3}$ is equivalent to $SU(2),$ we can write $S^{2} = \frac{SU(2)}{U(1)}$ so that $S^3$ can be constructed from $S^{2}$ by Hopf fibration. The Abelian field $U(1)$ corresponds to monopole field, which gives rise to the spin of the system. The magnetic flux line associated with this represents the vortex line. The vortex in an EVB corresponds to the spin vortex. Thus the introduction of the vortex line, which gives rise to the spin degrees of freedom extends the localization region from $S^{2}$ to $S^{3}$ and incorporates an angle $\chi$ corresponding to the orientation around the vortex line. The situation is exhibited clearly in the Skyrmion model of an electron, where it is considered that a scalar electron rotates around a direction vector(vortex line), which gives rise to the spin degrees of freedom \cite{14,24}. This leads to the introduction of a field function
of the form $\phi(x_{\mu},\xi_{\mu}),$ where $\vec{x}$ is the spatial coordinate and $\vec{\xi}$ is the 3-vector representing the direction vector. In this case the wave function takes into account the polar coordinates(r,$\theta,\phi$) for the spatial coordinate $\vec{x}$ and an angle $\chi$ to specify the rotational orientation around the direction vector $\vec{\xi}.$ The eigen-value of the operator $i\frac{\partial}{\partial \chi}$ represents the internal helicity \cite{24}. For an extended body represented by the de-Sitter group SO(4,1), $\theta$, $ \phi $ and $ \chi $ just represent the Euler angles. In 3 space dimensions these three Euler angles have their correspondence to an axisymmetric system, where the anisotropy is introduced along a particular direction and the components of the linear momentum satisfy a commutation relation of the form,
\beq [p_{i},p_{j}]=i\mu\epsilon_{ijk}\dfrac{x^{k}}{r^{3}}.\eeq Evidently $\mu$ represents the charge of a magnetic monopole.
The monopole harmonics incorporating the term $\mu$ has been extensively studied by Fierz \cite{Fierz}, Hurst \cite{Hurst} as well as by  Wu and Yang \cite{Wu}. Following them we can write
\begin{eqnarray} 
y_{l}^{m,\mu} &=& (1+x)^{\dfrac{(m-\mu)}{2}}(1-x)^{-\dfrac{(m+\mu)}{2}}\nonumber\\&& \dfrac{d^{l-m}}{d_{x}^{l-m}}\left[ (1+x)^{l-m}(1-x)^{l+\mu}\right]e^{im\phi}e^{-i\mu\chi},
\end{eqnarray}
where $x=cos\theta$. Here $l$ is the orbital angular momentum of a charged particle in the field of a magnetic monopole having charge $\mu$ and $m$ is the eigen value of the  $z$ - component of $\bf{L} $.

For specific values of  $l=\frac{1}{2},$ $m= \pm\frac{1}{2},$ $\mu =\pm \frac{1}{2},$ the above expression yields for $Y_{l}^{m,\mu}$
\begin{eqnarray}
Y_{\frac{1}{2}}^{\frac{1}{2},\frac{1}{2}} &=& sin\frac{\theta}{2}e^{i(\phi-\chi)/2}\nonumber\\
Y_{\frac{1}{2}}^{-\frac{1}{2},\frac{1}{2}} &=&cos\frac{\theta}{2}e^{-i(\phi+\chi)/2}\nonumber\\
Y_{\frac{1}{2}}^{\frac{1}{2},-\frac{1}{2}} &=& cos\frac{\theta}{2}e^{i(\phi+\chi)/2}\nonumber\\
Y_{\frac{1}{2}}^{-\frac{1}{2},-\frac{1}{2}} &=& sin\frac{\theta}{2}e^{-i(\phi-\chi)/2}
\end{eqnarray} 
​This represents monopole harmonics for half-integral OAM $l=\frac{1}{2}$​ with $\mu =\pm \frac{1}{2}.$  The doublet \begin{equation}
\phi  = \left( \begin{array}{c}
\phi_{1}\\
\phi_{2}
\end{array} \right),
\end{equation}
​with $\phi_{1} = Y_{\frac{1}{2}}^{\frac{1}{2},\frac{1}{2}}$​ and $\phi_{2} = Y_{\frac{1}{2}}^{-\frac{1}{2},\frac{1}{2}}$
​corresponds to a two component spinor. The charge conjugate state is given by
\begin{equation}
{\bar{\phi}}  = \left( \begin{array}{c}
{\bar{\phi}}_{1}\\
{\bar \phi}_{2}
\end{array} \right),
\end{equation}
​with ${\bar \phi}_{1} = Y_{\frac{1}{2}}^{-\frac{1}{2},-\frac{1}{2}}$​ and ${\bar \phi}_{2} = Y_{\frac{1}{2}}^{\frac{1}{2},-\frac{1}{2}}.$
This shows that a fermion can be viewed as a scalar particle moving with half-integral OAM in the field of a magnetic monopole and represents a skyrmion. This suggests that fractional OAM states have their relevance in the context of magnetic monopoles.

The expression (6) is valid when $\mu$ is quantized having values $0,~ \pm \frac{1}{2},~\pm 1.........$ so that we have integer and fractional OAM with half integer  value. However when $\mu$ is non-quantized, OAM can take any arbitrary fractional value. A quantum state with fractional OAM with non-quantized monopole charge is given by  $|M\rangle$ with $M=\ell+\mu$ where $\ell$ is an integer and $\mu $ is the fractional part lying between $0$ and $1$ . It may be mentioned that we have considered the states $|M\rangle$ such that it is applicable to the quantized value of $\mu = \frac{1}{2}$ also. The monopole harmonics contain the term involving $\mu$ in the form $e^{-i\mu\chi},$ where $\chi$ denotes the orientation  of the vortex line i.e the position of the discontinuity. The angle $\chi$ takes the value $0<\chi<2\pi.$ Generalizing this to the non-quantized values of $\mu,$ we note that the term containing the value of $\mu$ in the state$|M\rangle$ is of the form $e^{-i\mu\chi}.$  Indeed a generalised expression for the fractional OAM state $|M\rangle$ with $M=\ell+\mu$, $\ell(\mu)$ being an integer (fraction), can be derived  where these states are decomposed into the basis of integer OAM states. This is  given by \cite{7,8}   
\beq |M(\chi)\rangle=\sum_{\ell^{~'}} c_{\ell^{~'}}|M(\chi)\rangle|\ell\rangle \label{a},\eeq
where the coefficients $c_{\ell^{~'}}|M(\chi)\rangle$ are given by
\begin{widetext}
	\begin{eqnarray}	
	c_{\ell^{~'}}|M(\chi)\rangle &=& exp(-i\mu\chi)\frac{i exp[(i(M-\ell^{~'})\theta_{0}]}{2\pi(M-\ell^{~'})}exp[i(M-\ell^{~'})\chi]\nonumber\\
	&& (1-exp(i\mu 2\pi))\label{b}.
	\end{eqnarray}
\end{widetext}
with $\ell^{'} \in Z $. 
It is noted that the expression (11) contains the term $e^{- i\mu \chi}$ as we have in the monopole harmonics given by the expression (6). In fact such a term arises here from the consideration of the 
single-valuedness criterion derived from the position of the branch cut $\chi$ from the multivalued function $e^{i\mu\theta_n}$ where $ \theta_n=\theta_0+\frac{2\pi n}{2\ell+1}$ varying from $-\ell, -\ell+1,....+\ell$ $\theta_0$ being the starting point of the interval $[\theta_0,\theta_0+2\pi]$. Thus we have the dependance of the fractional OAM states on the dislocation $\chi$.

A characteristic feature of the state $|M(\chi)\rangle $ is that the probabilities $P_{\ell^\prime}(M)$ of observing the fractional OAM state for specific values of $\ell^\prime$ given by the modulus square of the probability amplitudes are independent of the angle $\chi$ \cite{8}. In fact we have 
\beq
P_{l'}(M)=|c_{l'}M(\chi)|^2=\frac{\sin^2(\mu\pi)}{(M-l')^2\pi^2}
\eeq.
For $M=l\in Z $ and $l-l' \neq 0$, the fractional part $\mu=0$ so that we have vanishing OAM probability. For $l'$, we can write
\beq
P_{l'}(M)=\lim_{\mu\rightarrow 0} P_{\ell'}(l+\mu)= \frac{1}{2\pi^2} \lim_{\mu\rightarrow 0} \frac{1-cos(2\pi n)}{(l+\mu-l')^2}=\delta_{i i}
\eeq
This means that for integer $M$ the OAM distribution is singular and is equipped with a single non-vanishing probability at $M=l'$. For fractional values of $M$, the probabilities are peaked around the nearest integer to $M$.  

In case of light beams, in cylindrical coordinate the field amplitude of the Laguerre-Gaussian mode is given by
\begin{widetext}
\begin{eqnarray} u^{\ell}_{p}(\rho,\phi,z) &=& \frac{c_{\ell p}}{\sqrt{w(z)}}\left(\frac{\rho\sqrt{2}}{w(z)}\right)^{|\ell|}exp\left(-\frac{\rho^2}{w^{2}(z)}\right)L^{|\ell|}_{p}\left(\frac{2\rho^2}{w^{2}(z)}\right)exp\left(i\frac{\rho^2}{w^{2}(z)}\frac{z}{z_{R}}\right)exp(i\ell\phi)\nonumber\\
&&exp\left[-i(2p+|\ell|+1)tan^{-1}(\frac{z}{z_{R}})\right]\label{c},\end{eqnarray}
\end{widetext}
where $w(z)$ is the Gaussian spot size.

\beq w(z) = \sqrt{2\frac{z_{R}^{2}+z^{2}}{k z_{R}}} = w_{0}\sqrt{1+\frac{z^{2}}{ z_{R}^{2}}}. \label{d}\eeq Here $k$ denotes the wave number and $z_{R}$ the Rayleigh range, $w_{0}$ the beam waist and $L^{|\ell|}_{p}$ are the Laguerre polynomials.
The normalization constants are given by
\beq c_{\ell p} = \sqrt{\frac{2p!}{\pi(|\ell|+p)!}}.\label{e}\eeq

The Gouy phase  $exp\left[-i(2p+|\ell |+1)tan^{-1}(\frac{z}{z_{R}})\right]$ describes the phase change as the beam moves through the beam waist situated at $z = 0.$ As it has been observed that the finite superposition of the LG modes is centred around the nearest integer to $M$, we can write 
\beq \psi_{M(\chi)}(\rho,\phi,z) = \sum_{\ell^{'} = \ell^{}_{min}}^{\ell_{max}}c_{\ell^{'}}|M(\chi)|u^{\ell^{'}}_{p}(\rho,\phi,z) \eeq 
Indeed the distribution of the coefficients $|c_{\ell}|^{2}$ shows that modes with an OAM index $\ell^\prime$ very different from $M$ contribute only little to the superposition and for all practical purposes, we can take the finite superposition of the LG modes which is centred around the nearest integer to $M.$
It is noted here that for every mode the sum $(2p+|\ell|+1)$ is equal to $|\ell_{max}|+1$ or $\ell_{max}.$

It is to be mentioned that when the fractional OAM state is decomposed into the basis of integer OAM states, it is not possible to have the same Gouy phase for all modes. This follows from the fact that, the superposition involves odd and even values of $\ell,$ but $p$ has to be an integer. However we can restrict the number of Gouy phases to two, such that one is for even $\ell$ and the other is for odd $\ell.$ Indeed when OVBs  having fractional OAM states are generated from the superposition of LG modes, the instability of these states leading to the change upon propagation arises from the interference between the LG modes with different Gouy phases. By choosing proper radial indices $p,$ for each value of $\ell,$ the instability of the fractional OAM states can be suppressed \cite{7}.  
It is noted that when the fractional value of $\mu$ is associated with the monopole charge, the instability arises due to the variation of $\mu$ with length scale $L$ according to the RG flow equation. We can relate this length scale with the radial index $p$ in the Gouy phase so that the change in $L$ will induce fluctuation in the Gouy phase factor. Due to the constraint that the radial index $p$ must be an integer, it is not possible to have the same Gouy phase for all modes in the superposition. Thus we note that in OVBs the change in the monopole charge $\mu$ with the length scale effectively leads to the interference between the LG modes with different Gouy phases giving rise to the instability. A proper choice of $p$ for each value of $L$ helps to make the OVBs with fractional OAM states stable upon propagation.

\section{fractional OAM states and electron vortex beams in an external magnetic field}

Here we consider the 
fractional OAM states  in EVBs when these propagate in an external magnetic field. As mentioned above, fractional OAM states arise when we have tilted vortex associated with the non-quantized value of the monopole charge $\mu.$ The states become unstable upon propagation, since the RG flow suggests that $\mu$ changes with the length scale $L.$ However, stability of the beam can be attained when fractional OAM states are generated from EVBs propagating in a magnetic field. 
We consider that the magnetic field ${\bf B}(r)$ is generated by a monopole of strength $\alpha$ situated at the origin of a unit sphere. The flux passing through the surface $\Sigma$ is given by
\beq
\Phi|_{\Sigma}=\int_\Sigma \vec{B} d{\vec{S}} =2\pi \alpha
\eeq
This effectively corresponds to the Berry phase attained by a particle moving around the flux in a closed contour $C$ corresponding to the holonomy
\beq
\Phi_B=\int_C {\vec{A}} d{\vec{r}}
\eeq

If the magnetic field $B(r)$ is an axially symmetric longitudinal one, the vortex vector potential can be chosen as 
\beq \vec{A}({\vec r}) = \frac{B(r)r}{2}\hat{\vec{e}}_{\phi},\eeq where $\vec{B} = \vec{\nabla}\times\vec{A}.$  For the  magnetic flux $\phi$, we can write the vortex vector potential as  
\beq \vec{A}({\vec r}) = 
\frac{\phi}{2\pi r}\hat{\vec{e}}_{\phi} = \frac{\alpha}{r}\hat{\vec{e}}_{\phi},\eeq with  $\alpha = \frac{\phi}{2\pi}$ being denoted as the magnetic field parameter. 

It may be mentioned here that Bliokh et al [24], considered EVBs in the presence of a single magnetic flux line corresponding to an infinitely thin shielded solenoid directed along the $z$ axis and containing the flux $\phi.$ The situation involves the vortex vector potential given by $\vec{A}(\vec{r})=\frac{\phi}{2\pi r}\vec{e}_{a} = \frac{\alpha}{r}\vec{e}_{a},$ where $\alpha$ represents the vortex charge of the vector potential $\vec{A}.$ This gives rise to Bessel beams in contrast to the LG beams. However the Bessel beams here are characterized by the fact that the Bessel function order is shifted by $\alpha,$ so that the wave function involves $J_{l-\alpha}(kr)$ instead of   $J_{l}(kr).$

Indeed the expectation value of OAM yields
\beq \langle L_{z}\rangle = (l-\alpha).\eeq 

This resembles the SOI but here it is caused by the Zeeman type interaction between OAM and magnetic field. However, this does not give rise to Zeeman energy since the wave function with  $ \langle L_{z}\rangle \neq 0$ is localized outside the area of the magnetic field. In fact the situation here involves Dirac phase  instead of the Berry phase when the magnetic field is associated with  monopole charge. 
The Dirac phase  is given by

\beq \phi|_{D} = \int_{C} \vec{A}. d\vec{r}= 2\pi\alpha,\eeq
​where $C$ is a closed loop characterized by $C$= ($r$=const. $ \phi\in [0,2\pi].$ 

It is noted that 
in the sharp point limit of the skyrmionic model of an electron, we can consider the non-relativistic situation. The Schroedinger equation in cylindrical coordinates for an electron in the magnetic field is given by 
\beq -\frac{\hbar^{2}}{2m}\left[\frac{1}{r}\frac{\partial}{\partial r}\left(r\frac{\partial}{\partial r}\right) + \frac{1}{r^{2}}\left(\frac{\partial}{\partial \phi} + ig\frac{2r^{2}}{w_{m}^{2}}\right)^{2} + \frac{\partial^{2}}{\partial z^{2}}\right]\psi = E\psi,\label{4}\eeq
where $w_{m} = 2\sqrt{\frac{\hbar}{|eB|}}$ is the magnetic length parameter, $g = sgn B = \pm 1$ indicates the direction of the magnetic field. The parameter $\frac{2r^{2}}{w_{m}^{2}}$  here essentially represents the monopole charge which follows from (20) and (21). The solution of eqn (\ref{4}) has the form of non-diffracting Laguerre-Gaussian (LG) beams given by \cite{28}
\beq \psi^{L}_{\ell,p} \simeq \left(\frac{r}{w_{m}}\right)^{|\ell|}L_{p}^{|\ell|}\left(\frac{2r^{2}}{w_{m}^{2}}\right)exp\left(-\frac{r^{2}}{w_{m}^{2}}\right)exp\left[i(\ell\phi+k_{z}z)\right].\label{6}\eeq
 It is noted that Landau LG modes yields exact solution. The wave numbers satisfy the dispersion  relation 
\beq E = \frac{k^{2}}{2m}-\Omega \ell +|\Omega| (2p+|\ell|+1).\eeq
where  $\Omega = \frac{eB}{2m}$ is the Larmor frequency. Here $E_{||} = \frac{k^{2}}{2m}$ is the energy of the free longitudinal motion. The transverse motion energy is given by $E_{\perp} = E_{Z}+ E_{g},$ where $E_{Z} = \Omega \ell$ represents the Zeeman energy and $E_{g} = |\Omega| (2p+|\ell|+1),$ $p$ being the radial quantum number, is associated with the Gouy phase of the diffractive LG modes. For the Landau state, the expected value of OAM corresponds to
\beq \langle L_{z}\rangle = \ell+g \langle\frac{2r^{2}}{\omega_{m}^{2}}\rangle,\eeq where $g=\pm 1$ is the sign factor and $\langle\frac{2r^{2}}{\omega_{m}^{2}}\rangle$ is given by
\beq \langle\frac{2r^{2}}{\omega_{m}^{2}}\rangle = \frac{\langle\psi |\frac{2r^{2}}{\omega_{m}^{2}}|\psi\rangle}{\langle\psi|\psi\rangle} =  (2p+|\ell|+1).\eeq
​This determines the squared spot size of the LG beams. From this we have the relation 
 \beq \langle \alpha\rangle = \langle \frac{2r^{2}}{w_{m}^{2}}\rangle = 2p + |\ell| + 1,\label{gou}\eeq
where  $p=0,1,2....$ is the radial quantum number and $|\ell|$ is the azimuthal quantum number. So fractional OAM states given by $\ell+\mu$ where $\mu$ corresponds to the fractional value $\mu \in [0,1]$​ representing the monopole charge in free case is now modified as
\beq \mu_{eff} = \mu+ g(2p+|\ell|+1)\eeq
As it is mentioned, the instability in the fractional OAM states arises due to the change of $\mu$ upon propagation as the RG flow implies that the non-quantized monopole charge changes with the length scale. However, an appropriate choice of the radial quantum number $p$ essentially fixes a length scale, so that in this case these states become stable. Indeed when we use the modified monopole charge 
the state​​ $|M(\chi)\rangle$ will be transformed to $|​​\tilde M(\chi)\rangle​,​$​ where $​\tilde M(\chi) = \ell+\mu_{eff}.$ When ${\tilde{M}}(\chi)$ is decomposed into the basis of integer OAM states, from eqn.(12)​, we find that the probability $P_{\ell}({​\tilde M})$ is now changed to
\begin{eqnarray} P_{\ell}({​\tilde M}) &=& |c_{\ell'}[{​\tilde M}(\alpha)]^{2}|= \frac{sin^{2}[\mu_{eff} \pi]}{({​\tilde M}-\ell')^{2}\pi^{2}} \nonumber\\ &=& \frac{sin^{2}[\mu+ (2p+|\ell|+1))\pi]}{(l+\mu+ (2p+|\ell|+1)-\ell')^{2}\pi^{2}} .\end{eqnarray}
​Now noting that $sin^{2}[(N+\mu) \pi] = sin^{2}[\mu\pi]$,​ when $N$ is an integer. We finally can write,
\beq P_{\ell}({​\tilde M}) = \frac{sin^{2}[\mu\pi]}{(\ell+\mu+ (2p+|\ell|+1)-\ell')^{2}\pi^{2}} .\eeq
​As is evident from the above eqn.​, this probability $P_{\ell}({​\tilde M})$ vanishes for integer OAM states $(\mu=0)$. From eqn. (32)  we note that for various values of $p,$ $P_{l}({​\tilde M})$ changes and as $p$ increases this approaches towards zero. Thus with increasing values of $p$ we have the situation such that the contribution of the fractional part to the state ​ $|{​\tilde M}(\chi)\rangle$ gradually vanishes and it approaches towards an integer OAM state, leading to a stable configuration. Thus with the appropriate choice of $p,$ we can achieve stability. Thus we find that when,
EVBs move in an external magnetic field, the change in $\mu$ as depicted in eqn. (30), effectively makes the corresponding fractional OAM states stable upon propagation, when a proper radial index $p$ is chosen. 

It is to be noted that, unlike OVBs, here we need not require a superposition of LG modes to have  stable fractional OAM state. This suggests that, for EVBs moving in an axisymmetric longitudinal magnetic field, we can have stable fractional OAM states. In Fig. 1, we have plotted the probability $P({\tilde{M}})$ for various values of $\mu$ and $p$ in (a) with $\ell=2,\ell^\prime=1$ and in (b) with $\ell=-2,\ell^\prime=1$. 

As the Gouy phase factor $2p+|\ell|+1,$ essentially corresponds to the expectation value of the magnetic monopole charge,  the stability of these states can be achieved by properly tuning the magnetic field. This makes the situation of EVBs different from OVBs having stable fractional OAM states.

\begin{figure}\label{fig}
	\includegraphics[width=6 cm]{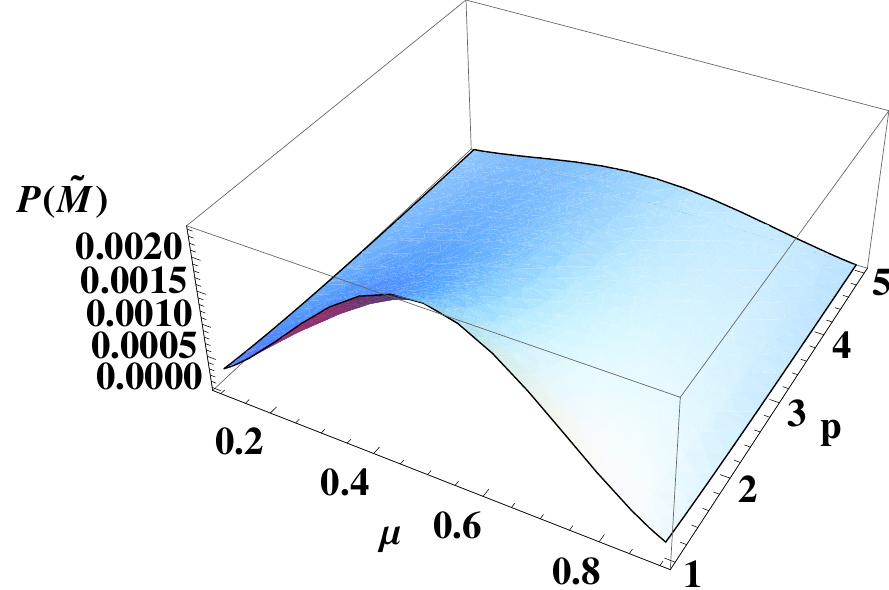}
	\includegraphics[width=6 cm]{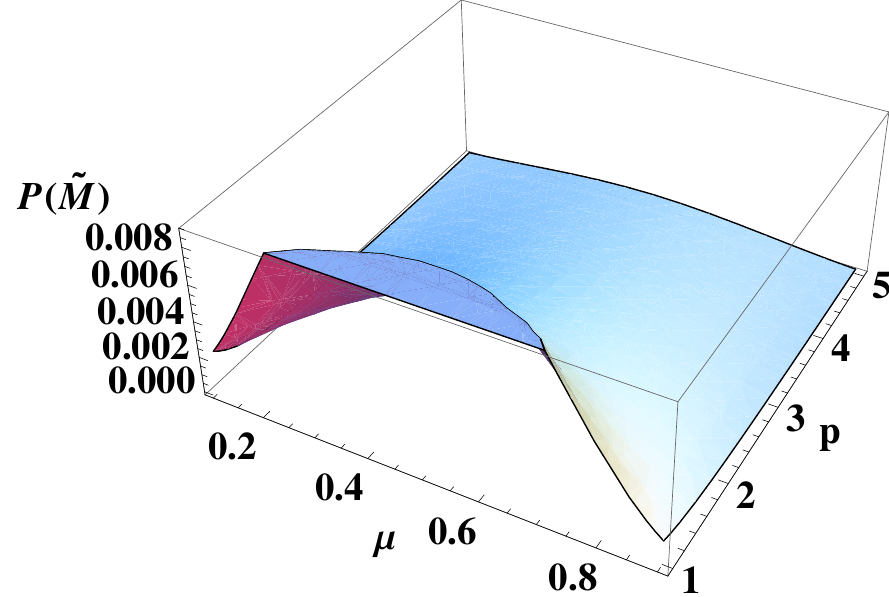}	
	\caption{ (Color online): The Probability $P(\tilde{M})$ for various values of $\mu$ and $p$ for   (a) $\ell=2,~\ell'=1$ and  (b)$\ell=-2,~\ell'=1$
		~~Down:(b) . }
\end{figure}


​

\section{Discussion} 
The EVBs carrying fractional OAM have the specific feature that the geometric phase acquired by the scalar electron orbiting around the vortex line involves non-quantized monopole charge. Fractional OAM states arise due to the spin-orbit coupling associated with the generation of the geometric phase. The RG flow of the monopole charge induces length scale dependence in it, which makes the EVBs unstable on propagation. However, when EVBs propagate in an external magnetic field, the change in the Berry phase factor due to incorporation of the Gouy phase makes the fractional OAM states stable, with a proper choice of the radial index. This makes the situation different from OVBs, where fractional OAM states become stable when these are generated by the synthesis of the LG modes with a minimal number of different Gouy phases. 

We have already pointed out in an earlier paper \cite{13}  that the temporal variation of the monopole charge in an EVB in free space leads to spin Hall effect. This arises from the anomalous velocity caused by the Berry curvature. In the presence of an external magnetic field, the modified Berry curvature leads to spin filtering such that either positive or negative spin states emerge in spin Hall currents with clustering of spin $1/2$ states \cite{18}.

To consider the vortex structure of beams with fractional OAM states we note that we have considered the vortex here to be a spin vortex, so the system of vortices in vortex beams may be viewed as a spin system. When the two adjacent spins (vortices) have opposite orientations we have a pair of entangled spins (vortices) . A tilted vortex represents a uniaxial anisotropy axis which can have various orientations when the wave front propagation direction is the z axis. As the vortices are tilted in nature, an entangled pair of alternating vortices corresponds to a hairpin configuration implying that the two vortices converge at a common turning point. Thus the vortex structure in EVBs will correspond to a chain of the pairs of alternating vortices leading to hairpin configuration as observed in OVBs \cite{mv,leach}.

It may be noted that fractional OAM states in OVBs and EVBs occur only in  non-paraxial beams.  Gotte et al \cite{8} have shown that by using a decomposition of Bessel beams rather than LG modes in OVBs, we can achieve a paraxial solution. It is observed that the paraxial solution approaches the exact solution for $z\rightarrow \infty$. However, it does not show the formation of the chain of vortices.

Finally, we may add here that fractional OAM state have already been studied in entangled photon pairs \cite{30,31}. Indeed as fractional OAM states involve non-quantized monopole charge, we have to take into account the effect of the Dirac string. However the observability of the Dirac string can be avoided, when we take into account an entangled state \cite{17}. In view of this we note that just like OVBs it is expected that, these states can also be exhibited for entangled electron pairs. A stable fractional OAM state thus become potentially important in the study of quantum information and foundation. \\


\begin{thebibliography}{999}
\bibitem{1} J.F. Nye and M.V. Berry, Proc R Soc. (London) A {\bf 336}, 165 (1974)  
\bibitem{allen}  L. Allen,  M. W. Beijersbergen , R. J. C. Spreeuw and   J. P. Woerdman, Phys. Rev. A {\bf 45}, 8185, (1992).
\bibitem{2} M. W. Beijersbergen, R. P. C. Coerwinkel, M. Kristensen and J. P. Woerdman, Optics Communications,  {\bf 112},321, (1994).
\bibitem{3}V. Yu. Bazhenov, M. V. Vasnetsov, M. S. Soskin, JETP Letters, {\bf 52}, 429, (1990).
\bibitem{opticslett} N. R. Heckenberg, R. M. Duff, C.P. Smith and A.G. White, Opt. Lett. {\bf 17}, 221 (1992)
\bibitem{7}   J.B.Gotte et al., J.Mod.Opt. {\bf 54}, 1723, (2006);Optics Express {\bf 16}, 993, (2008).
\bibitem{9}M. Uchida and A. Tonomura, Nature {\bf 464}, 737, (2010).
\bibitem{10} J. Verbeeck, H.Tian and P. Schattschneider,Nature {\bf 467}, 301, (2010).
\bibitem{11} B. J. McMorran et. al,  Science, {\bf 331}, 192, (2011).
\bibitem{12}K. Y. Bliokh et. al., Phys.Rev.Lett, {\bf 107}, 174802, (2011).
\bibitem{13}P. Bandyopadhyay, B. Basu, D. Chowdhury, Proc.R.Soc.(London) A {\bf 470}, 20130525, (2014).
\bibitem{16} P. Bandyopadhyay, B. Basu, D. Chowdhury,  Phys. Rev. Lett. {\bf 116}, 144801,(2016); Phys. Rev. Lett. {\bf 115}, 194801,(2015).
\bibitem{18}D. Chowdhury, B. Basu, P. Bandyopadhyay, Phys. Rev. A {\bf 91}, 033812 (2015).
\bibitem{14}P. Bandyopadhyay , K. Hajra, J. Math. Phys. {\bf 28}, 711, (1987).
\bibitem{24}D. Banerjee, , P. Bandyopadhyay, J. Math. Phys. {\bf 33}, 990, (1992).
\bibitem{27}P. Bandyopadhyay, Int. J. Mod Phys. A {\bf 15}, 1415, (2000).
\bibitem{17} P. Bandyopadhyay,  Proc. R. Soc.(London) A. {\bf 467}, 427, (2011).
\bibitem{Bayen} F  Bayen  and  J.  Niederlie, Czech.  J.  Phys.  B
{\bf 31},  1317,(1981).
\bibitem{Sir} P. Bandyopadhyay, Int. J. Theo.Phys. {\bf 26}, 131, (1987).
\bibitem{Fierz} M. Fierz,Helv.Phys.Acta,
{\bf 17},27,(1944).
\bibitem{Hurst}C.A Hurst, Ann. Phys {\bf 50},51, (1968).
\bibitem{Wu}T. T. Wu, C. N. Yang, Nuclear phys B, {\bf 107}, 365, (1976).
\bibitem{8} J.B.Gotte et al., J. Mod. Opt, {\bf 54}, 1723, (2007).
\bibitem{28}K. Y. Bliokh, P. Schattschneider, J. Verbeeck, and F. Nori, Phys. Rev. X {\bf 2}, 041011, (2012).
\bibitem{mv} M V Berry, J Opt. A {\bf 6}, 259 (2004)
\bibitem{leach} J. Leach, E. Yao and M. J Padgett, New Journal of Physics, {\bf 6}, 71, (2004)
\bibitem{30}S. S. R. Oemrawsingh, A. Aiello, E. R. Eliel, G. Nienhuis, and J. P. Woerdman, Phys. Rev. Lett. {\bf 92}, 217901, (2004).

 
\bibitem{31}G. F. Calvo, A. Picón, and A. Bramon, Phys. Rev. A {\bf 75}, 012319, (2007).
\end{thebibliography}
\end{document}